\documentclass[a4paper,11pt]{article}

\usepackage{pos}
\usepackage{lipsum} 

\title{A time variability test for neutrino sources identified by IceCube}

\ShortTitle{Time variability of neutrino sources with IceCube}

\author{The IceCube Collaboration \\{\normalsize \normalfont(a complete list of authors can be found at the end of the proceedings)}\\}

\emailAdd{pdave@gatech.edu}
\emailAdd{itaboada@gatech.edu}

\abstract{
IceCube has reported evidence for neutrino emission from the Seyfert-II galaxy NGC 1068 and the blazar TXS 0506+056. The former was identified in a time-integrated search, and the latter using time-dependent and multi-messenger methods. A natural question is: are sources identified in time-integrated searches consistent with a steady neutrino source? We present a non-parametric method, TAUNTON, to answer this question. Motivated by the Cram\'er-von Mises test, TAUNTON is an unbinned single-hypothesis method to identify deviations in neutrino data from the steady hypothesis. An advantage of TAUNTON is that it is sensitive to arbitrary deviations from the steady hypothesis. Here we present results of TAUNTON applied to a 8.7 year data-set of muon neutrino track events; the same data used to identify NGC 1068 at 4.2$\sigma$. We use TAUNTON on 51 objects, a subset (with $>$4 signal neutrinos) of the 110 objects studied in the NGC 1068 publication. We set a threshold of $3\sigma$ pre-trial to identify sources inconsistent with the steady hypothesis. TAUNTON reports a p-value of 0.9 for NGC 1068, consistent with the steady hypothesis. Using the time integrated fit, data for TXS~0506+056 is consistent with the steady hypothesis at $1.7\sigma$. Time variability is not identified for TXS~0506+056 because there are few neutrino events. 

\vspace{4mm}
{\bfseries Corresponding authors:}
Pranav Dave$^{1}$, Ignacio Taboada$^{1*}$\\
{$^{1}$ \itshape School of Physics and Center for Relativistic Astrophysics. Georgia Institute of Technology. Atlanta, GA. USA.}\\[4mm]
$^*$ Presenter

\ConferenceLogo{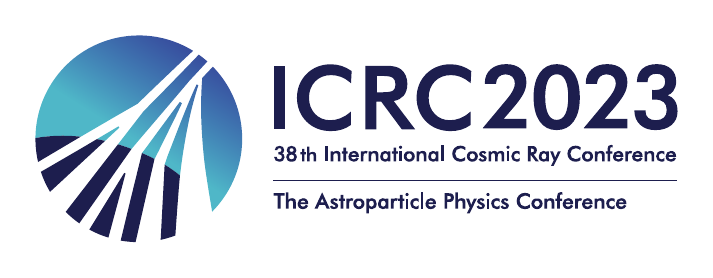}

\FullConference{The 38th International Cosmic Ray Conference (ICRC2023)\\ 26 July -- 3 August, 2023\\ Nagoya, Japan}
}

\begin{document}

\maketitle
\section{Introduction}\label{sec1}

IceCube, a cubic-kilometer neutrino detector operating at the South Pole, has discovered an extragalactic flux of neutrinos \cite{2014PhRvL.113j1101A}. The class of sources responsible for this flux have not been identified. Two candidate sources have been identified: the blazar TXS~0506+056 \cite{2018Sci...361.1378I,2018Sci...361..147I} and the Seyfert-II and starburst NGC~1068 \cite{2022Sci...378..538I}. Though both have Active Galactic Nuclei (AGN), these two sources have very different properties: TXS~0506+056 is distant, $d$=1.79~Gpc ($z=0.3365$ \cite{2018ApJ...854L..32P}), and NGC 1068 is nearby, $d = 14.4$~Mpc; the spectrum of TXS~0506+056 is hard, $\gamma=$2.0, and the spectrum of NGC 1068 is soft, $\gamma=$3.2. TXS~0506+056 is time variable in neutrinos, while previous studies and the results we present here indicate that NGC~1068 is steady.

The characterization of the time-dependence of neutrino sources is a fundamental objective of IceCube. Time variability can be used to understand the mechanisms responsible for neutrino production. AGN of all types can be highly variable across the entire electromagnetic spectrum. IceCube has previously used likelihood ratio methods to search for single or multiple neutrino flares from astrophysical objects. The likelihood ratio method is, as a virtue of the Neyman-Pearson lemma, the most powerful method to identify a flaring object. However the likelihood ratio method depends on PDFs being known perfectly, notably the time-PDF. The likelihood ratio method for single or multiple flares may fail to identify other time-dependent signals. In this work we take a different approach; the objective is to perform a single hypothesis test that compares data to what is expected for a time-steady source hypothesis. This approach has the advantage that arbitrary time-deviations from the steady hypothesis can be considered, not only flares. In this work we present an update of TAUNTON \cite{2021arXiv211006294D}, a non-parametric and unbinned method that tests the compatibility of IceCube data with the a steady neutrino source hypothesis. TAUNTON's objective is not to discover flaring sources, but to characterize the time variability of already identified neutrino candidate sources found in a time-integrated study. We present the application of the updated method to NGC~1068, TXS~0506+056 and another 49 objects that were studied in the time-integrated publication that reported NGC 1068 at 4.2$\sigma$ \cite{2022Sci...378..538I}.

TXS~0506+056 was identified via time-dependent methods. On September 22, 2017, a high-energy neutrino, IceCube-170922A, was publicly reported \cite{2017GCN.21916....1K}. Follow up observations revealed that this blazar was flaring in $\gamma$-rays. An accidental coincidence was ruled out at $3\sigma$ \cite{2018Sci...361.1378I}. A study of archival IceCube data identified an additional neutrino flare at $3.5\sigma$ that lasted for $\sim$5 months in 2014-2015 \cite{2018Sci...361..147I}. 

NGC~1068 was first identified as a 2.9$\sigma$ hot-spot using time-integrated methods \cite{2020PhRvL.124e1103A}. 
A search for single or multiple flares did not reveal activity by NGC~1068 \cite{2021ApJ...920L..45A}. In ICRC~2021 we presented TAUNTON results for NGC~1068, using the 2020 study, and found data in the direction of NGC~1068 to be consistent with a steady hypothesis \cite{2021arXiv211006294D}. The most exciting development on the observation of NGC~1068 with neutrinos was reported in 2022. This included data reprocessing, known as pass2, and several improvements on reconstructions and data analysis, which resulted in a post-trial significance of 4.2$\sigma$ \cite{2022Sci...378..538I}. NGC~1068 is reported in this latter study to have an excess of 79 neutrinos and a spectral index of $\gamma=$3.2. Also included in this latter work is a time-integrated fit of TXS~0506+056, which can be described as an excess of 5 neutrinos and spectral index of $\gamma=2.0$. It is expect that, for TXS~0506+056, time-integrated methods identify a different number of excess neutrinos with respect to methods that are explicitly time-dependent. A re-evaluation of TXS~0506+056 using time-dependent methods, pass2 and improved reconstruction and analysis methods is also presented in this conference \cite{ICRC:NS10}. For all sources considered here, TAUNTON uses the outcome of the time-integrated study that resulted in the identification of NGC~1068 at 4.2$\sigma$.

\section{Method description}

\begin{figure}
\centering
\includegraphics[width=0.5\textwidth]{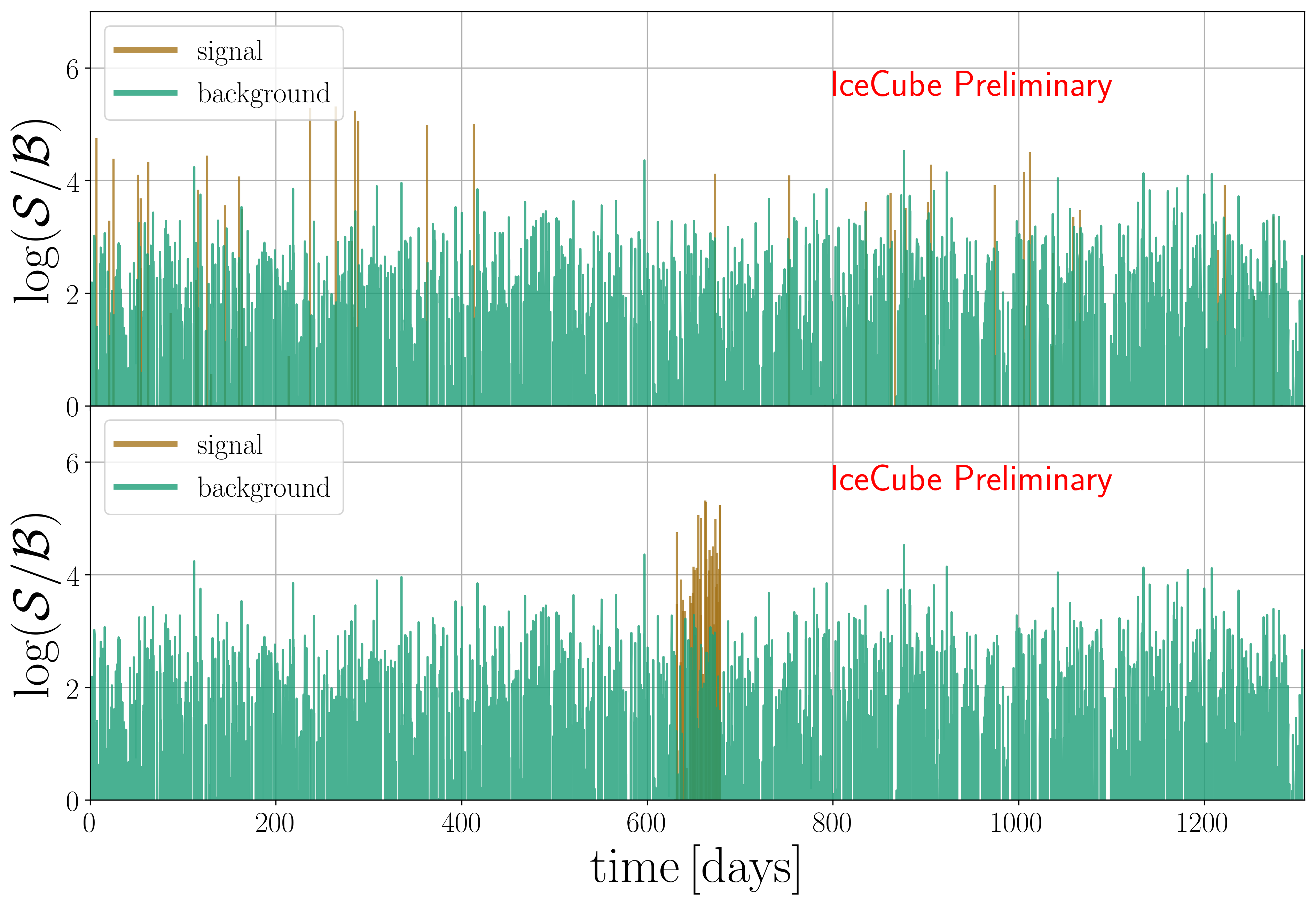}
\caption{\label{fig:timeseries}
Two example neutrino-curves. Both panels show background in green, obtained via scramblings. Orange indicates signal simulations. The top panel shows a simulated steady source, and the bottom panel shows a simulated top-hat flaring neutrino source. The vertical scale indicates the spatial $\mathcal{S/B}$ for each event. A minimum value of $\mathcal{S/B}$=1 has been used for this plot, which is why the green data (background from scramblings) is very dense.}
\end{figure}

TAUNTON \cite{2021arXiv211006294D} is based on the Cram\'er-von Mises test. This is a non-parametric, unbinned, single hypothesis test. In this test, the cumulative distribution function (CDF) of the hypothesis is compared to the empirical cumulative distribution function (EDF) of the observations. For TAUNTON the single hypothesis to be tested is that of a steady neutrino source. The CDF and EDF are calculated using the time difference between consecutive neutrino events. Figure \ref{fig:timeseries} shows two example neutrino time series. The top panel corresponds to a steady neutrino source, and the bottom panel corresponds to a flaring neutrino source. 

Let's start with a time-ordered set of $n+1$ neutrinos. Then $n$ time differences, $\Delta t_i = t_{i+1} - t_i$ can be calculated. The EDF of the time differences is:
\begin{equation}
F_n(\Delta t) = \frac{1}{n} \sum_{i=1}^n {w_i}_{\{ \Delta t_i < \Delta t\}},
\end{equation}
where $w_i$ is a weight, described later, assigned to each $\Delta t_i$. See that $F_n(\Delta t)$ increases in steps of value $w_i$ at exactly $\Delta t = \Delta_i$. That is, there is no binning in the construction of the EDF. Note that in the standard Cram\'er-von Mises test $w_i = 1$.

Neutrinos in the directional vicinity of a neutrino source may be due to background, notably atmospheric neutrinos, or due to true astrophysical signal. We calculate the signal to background ratio, $\mathcal{S}/\mathcal{B}$, for each neutrino based on the angular separation between the neutrino source and the reconstructed neutrino direction as well as the angular uncertainty of the neutrino. This per-neutrino $\mathcal{S}/\mathcal{B}$ is calculated using the exact same definition as with the likelihood-ratio methods used in time-integrated neutrino point source search methods \cite{2022Sci...378..538I}. For the NGC~1068 publication that identified it at 4.2$\sigma$, the angular uncertainty of muons - product of $\nu_\mu$ interactions with matter - are described using Kernel Density Estimations (KDE), which is a significant improvement over the prior assumption of a Rayleigh distribution. The use of KDEs has a significant impact on $\mathcal{S}$ and for events in the vicinity of a candidate neutrino source. 

We define the weight as $w_i = \sqrt{\log( \mathcal{S}_{i+1}/\mathcal{B}_{i+1}) \times \log( \mathcal{S}_{i}/\mathcal{B}_{i})}$, that is, both events $i$ and $i+1$ contribute to the weight $w_i$ assigned to time difference $\Delta t_i$. For mathematical convenience, the weights are normalized so that $\sum_{i=1}^n w_i = 1$ . The definition of weights is empirical; we tested multiple options to obtain the best performance \cite{2021arXiv211006294D}. 

\begin{figure}
\centering
\includegraphics[width=0.5\textwidth]{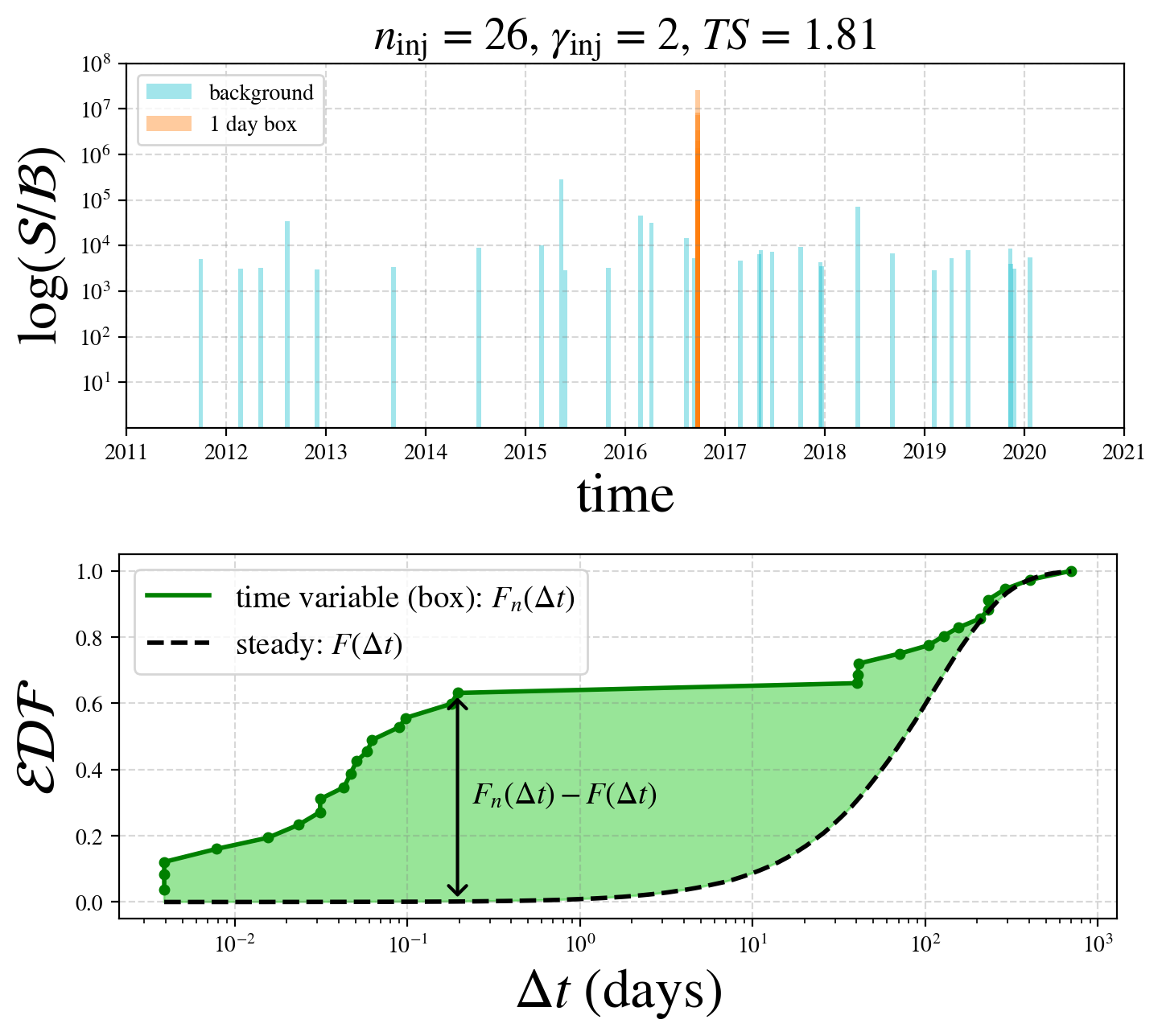}
\caption{\label{fig:exampleTAUNTON} The top panel shows the time curve of a 1 day long, strong neutrino flare. Signal injection is 26 events, and the spectral index is $\gamma=$-2.0. Events from background scramblings is shown in cyan; signal injections are shown in orange. The bottom panel shows the EDF, in solid green, $F_n(\Delta t)$ from the pseudo-experiment in the top panel as well as the cdf, in dashed black, $F(\Delta t)$ for the steady signal hypothesis. The cdf has also been constructed with 26 injected signal events and $\gamma=$-2.0. The separation between the EDF and the cdf, $F(\Delta t) - F_(\Delta t)$ is fundamental to TAUNTON. To illustrate the process of constructing the EDF for data and cdf for the hypothesis, a high $\mathcal{S/B}=10^3$ has been used on the top panel. Applying this example to TAUNTON results in a test statistic of $TS=1.81$.
}
\end{figure}

Because IceCube is at the geographic South Pole, a randomization - or scrambling -  of the time of an event, while keeping detector coordinates that describe the event constant, is equivalent to randomizing the event's right ascension. While the uptime of IceCube, is excellent, $>$95\%, scramblings randomize times exclude those periods when IceCube is not collecting data. Any given scrambling reproduces all the relevant characteristics of background, which includes atmospheric neutrinos, the all-sky extragalactic neutrino flux and (the very small contribution of) mis-reconstructed down-going cosmic ray muons. A simulated steady neutrino source can be injected on top of a given scrambling. Therefore a combination of scramblings and signal simulation can produce background-like data as well as signal plus background-like data. Let's call both cases pseudo-experiments.  

Signal plus background pseudo-experiments can be produced many times and the EDF for each instance is calculated. The hypothesis cdf is then calculated by averaging over the ensemble of signal plus background EDFs. The top panel of figure \ref{fig:exampleTAUNTON} shows an example neutrino time series for an extremely narrow flare. The bottom panels shows the EDF for this time series and the cdf for a steady hypothesis with the same number of signal neutrinos. 

Following the Cram\'er-von Mises test, we define a test statistic, $TS$, as:
\begin{equation}
TS = \sqrt{ N_{ev} \int_0^1 (F(\Delta t) - F_n(\Delta t))^2 dF },
\label{eq:TAUNTON}
\end{equation}
where $N_{ev}$ is the number of events considered for this test, to be defined later. The test statistic distribution for steady neutrino sources pseudo-experiments, can be used to define a p-value for data. For a simulated steady source, the test statistic distribution depends on the number of signal events, the declination of the source and the spectral index of the source. 

We adopt a per-source TAUNTON significance of 3$\sigma$ as the threshold to identify non-steady neutrino sources. We adopt this threshold because TAUNTON is not a discovery tool, but a characterization tool. Also, we don't consider trials on testing multiple sources, as we want to ask for each candidate source if it is consistent with the steady source hypothesis. 

To identify the events, $N_{ev}$ that are used on eqn.\ref{eq:TAUNTON}, we begin by selecting a $5^\circ \times 5^\circ$ region around the source being studied. We then select the events with highest spatial $\mathcal{S/B}$. We studied the optimal value of events, $N_{ev}$ to select, and found it to be identical to the signal strenght injected, $N_{inj}$. In real data, $N_{inj}$ is, of course, not known, so we use the best fit signal number, $N_{fit}$. Unlike our prior work on TAUNTON \cite{2021arXiv211006294D}, a benefit of the methods used in the 4.2$\sigma$ observation of NGC~1068 is that there is little bias between $N_{inj}$ and $N_{fit}$. Nevertheless we use signal injections to find the median value of $N_{inj}$ that results on an average $N_{fit}$. 
 
\section{Comparison to Fermi LAT's Time Variability Index}

\begin{figure}
\centering
\includegraphics[width=0.5\textwidth]{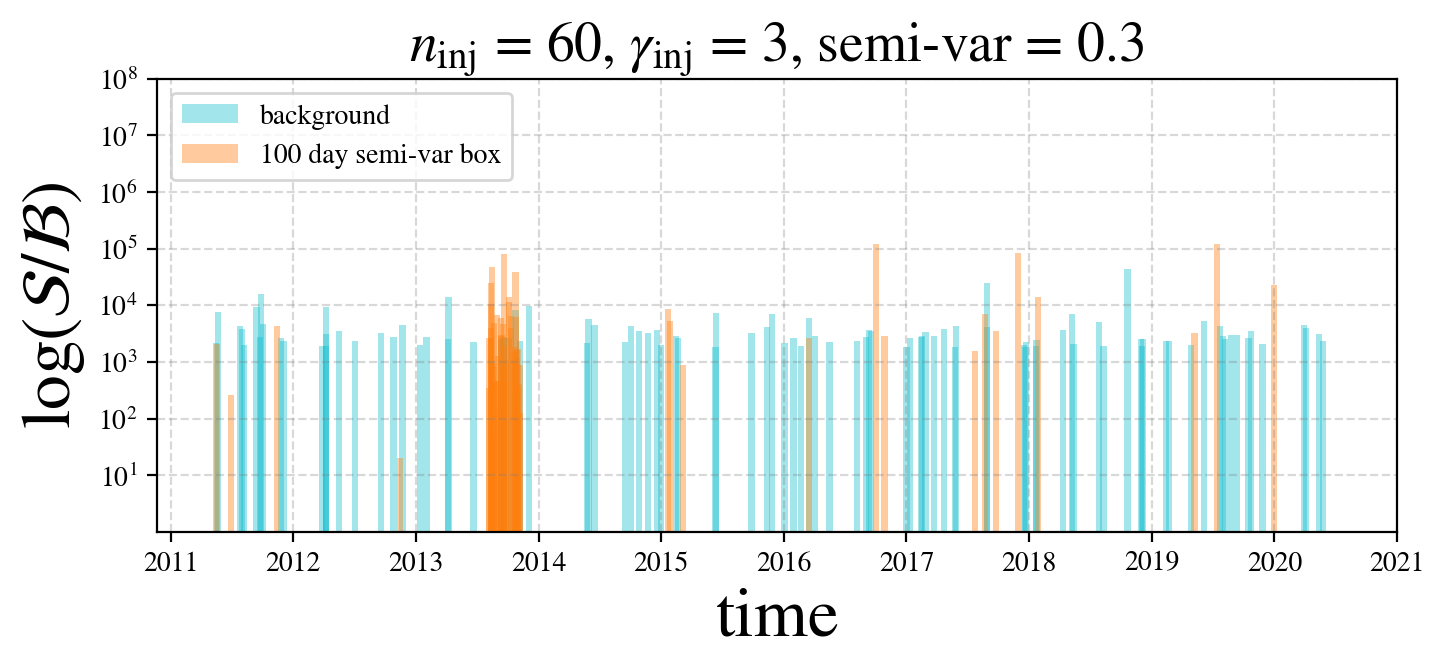}
\caption{\label{fig:semi} Example time-curve for a semi-variable neutrino pseudo-experiment. The point has a declination of $\delta=0^\circ$; 60 neutrinos have been injected with a spectral index of, $\gamma=3.0$. The semi-variable fraction is 0.3, that is, 30\% of the injected neutrinos are in the steady portion of the flux and 70\% are in the top-hat flare. The vertical scale shows $\mathcal{S/B}$ for all events. Scrambled background is shown in cyan, injected signal is shown in orange. For clarity, only $\mathcal{S/B}>10^3$ are shown.}
\end{figure}

The Fermi LAT collaboration has defined a  time variability index \cite{2020ApJS..247...33A}. Fermi LAT data is binned in 1 month intervals (2FGL) or 2 months / 1 year (4FGL). For each bin, the source spectrum is fitted, and the photon flux is calculated. In case emission is too weak, a limit on the photon flux is set. A binned-likelihood ratio is then used to compared all the time binned data to the expectation of a steady hypothesis. The test statistic of this likelihood ratio is used as a threshold to define time variability. 

We find that Fermi LAT's time variability index is not well adapted to IceCube. This is because, the most significant point source by IceCube is NGC~1068 at 4.2$\sigma$. Any time binning would result in fluxes that are too weak to be observable. The paucity of neutrino data is what makes us prefer an unbinned method, such as TAUNTON.

\section{Benchmarking the Sensitivity to Time Variability}

\begin{figure}
\centering
\includegraphics[width=0.5\textwidth]{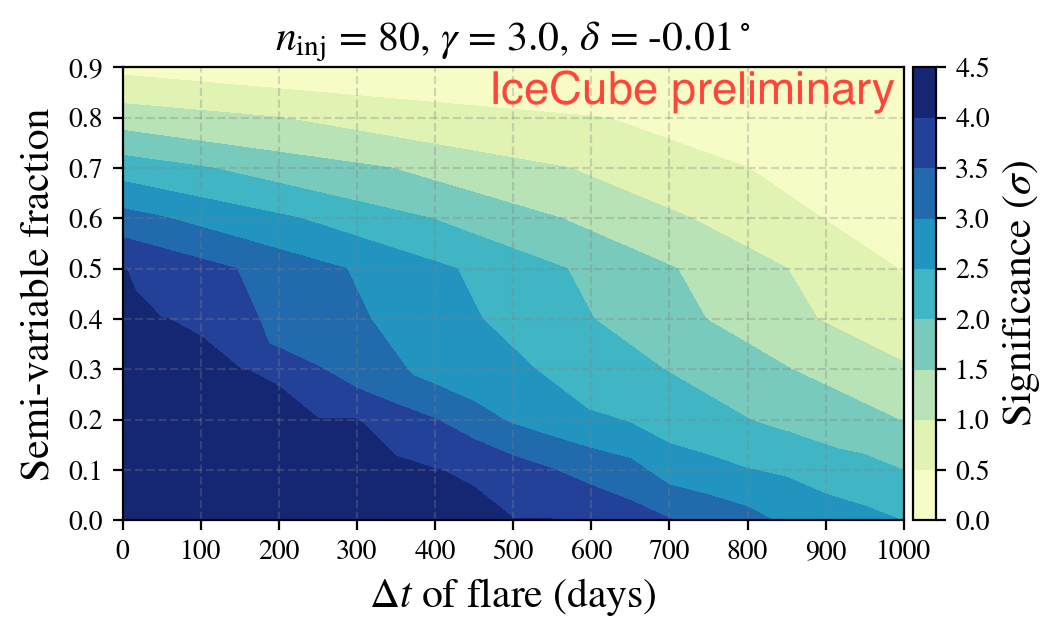}
\caption{\label{fig:semitest} TAUNTON significance for simulated semi-variable signal. A semi-variable fraction of 1.0 corresponds to a steady signal; a semi-varible fraction of 0.0 corresponds to a single top-hat flare. As defined in the text, we define sources as being inconsistent with a steady signal, if the TAUNTON significance is 3$\sigma$.}
\end{figure}

As described before, TAUNTON is able to identify time variability for arbitrary neutrino-curves. It is nevertheless interesting to explore the performance of TAUNTON for a specific simple case. We define here a \textit{semi-variable} signal to benchmark the performance of TAUNTON.

The semi-variable signal is a pseudo-experiment that combines a steady source and a single top-hat shaped flare. 
We call the ratio of the injected neutrinos in the steady component to all the injected neutrinos, the semi-variable fraction. 
A semi-variable fraction of 1.0, corresponds to a steady sources, while a fraction of 0.0 corresponds to a source that emits neutrinos in a single top-hat flare. Except in those extreme cases, neutrino signal can't be represented as a single flare, a combination of flares or a steady source \footnote{Actually, the steady component can always be represented as a top-hat flare that has the same width as the observation time window. But that is an intrinsic limitation of finite observation time!}. Figure \ref{fig:semi} shows an example semi-variable pseudo-experiment of 60 injected signal neutrinos with a spectral index of 3.0, a semi-variable fraction of 0.3 and a top-hat flare that lasts 100 days.  

We have tested a semi-variable signal at declination $\delta=0.0^\circ$, $N_{inj}=80$ and spectral index $\gamma=3.0$ for semi-variable fraction from 0.0 to 1.0 and for top-hat flares from 1 day to 1,000 days. We chose these examples, as this simulation resembles - but it is not identical - NGC~1068. Figure \ref{fig:semitest} shows the significance of deviation with respect to the steady hypothesis for these example cases. We find that, using the $3\sigma$ threshold defined above, a 1 day flare would be identified, on top of a steady signal, if the semi-variable fraction is $\lesssim$0.6, that is, the flare contains no more than 40\% of the injected signal. For flares of $\sim$1~year, TAUNTON can reject the steady hypothesis with a semi-variable fraction of $\lesssim$0.3, or that the flare contains no more than 70\% of the signal. In the most extreme case of a single flare and no steady signal component, i.e., a semi-variable fraction of 0.0, TAUNTON would be reject the steady hypothesis for flares as wide as $\sim$800~days.

\begin{figure}
\centering
\includegraphics[width=0.5\textwidth]{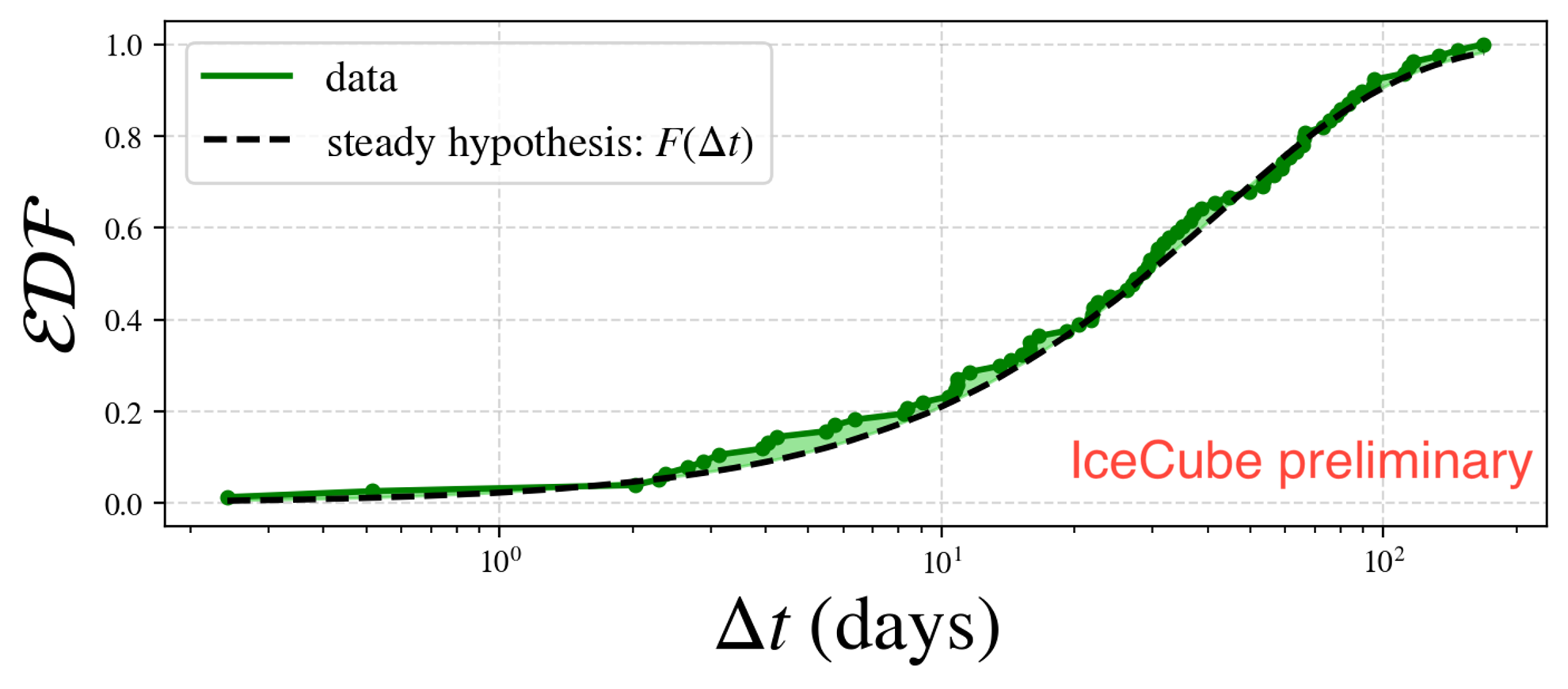}
\caption{\label{fig:ngc1068result} The black dashline shows the cdf for the steady hypothesis for NGC~1068. The green line shows the EDF.}
\end{figure}

\section{Results}

The NGC~1068 4.2$\sigma$ observation was based on a 110 source list. In order to construct a minimally meaningful EDF, we set a minimum of 
$N_{ev}=$4 events. Of the 110 sources, 51 meet this criteria, including NGC~1068 and TXS~0506+056. 
As mentioned before, in this time-integrated study, the significance of TXS~0506+056 is lower than 3$\sigma$.

No source, out of the 51 objects indicated, has a significance that meets the 3$\sigma$ time variability. Only NGC~1068 and TXS~0506+056 are candidate neutrino sources. The best fit values for signal for NGC~1068 are $N_{fit}=79$ and $\gamma=3.2$ and we find $N_{ev}=79$ as described before. Figure \ref{fig:ngc1068result} shows the cdf for the steady hypothesis, using $N_{ev}=79$, $\gamma=3.2$ and $\delta=-0.01^\circ$, as well as the EDF from the data for NGC 1068. The TAUNTON p-value for NGC~1068 is $p=0.9$, we thus find NGC~1068 data consistent with the steady hypothesis. 

\begin{figure}
\centering
\includegraphics[width=0.5\textwidth]{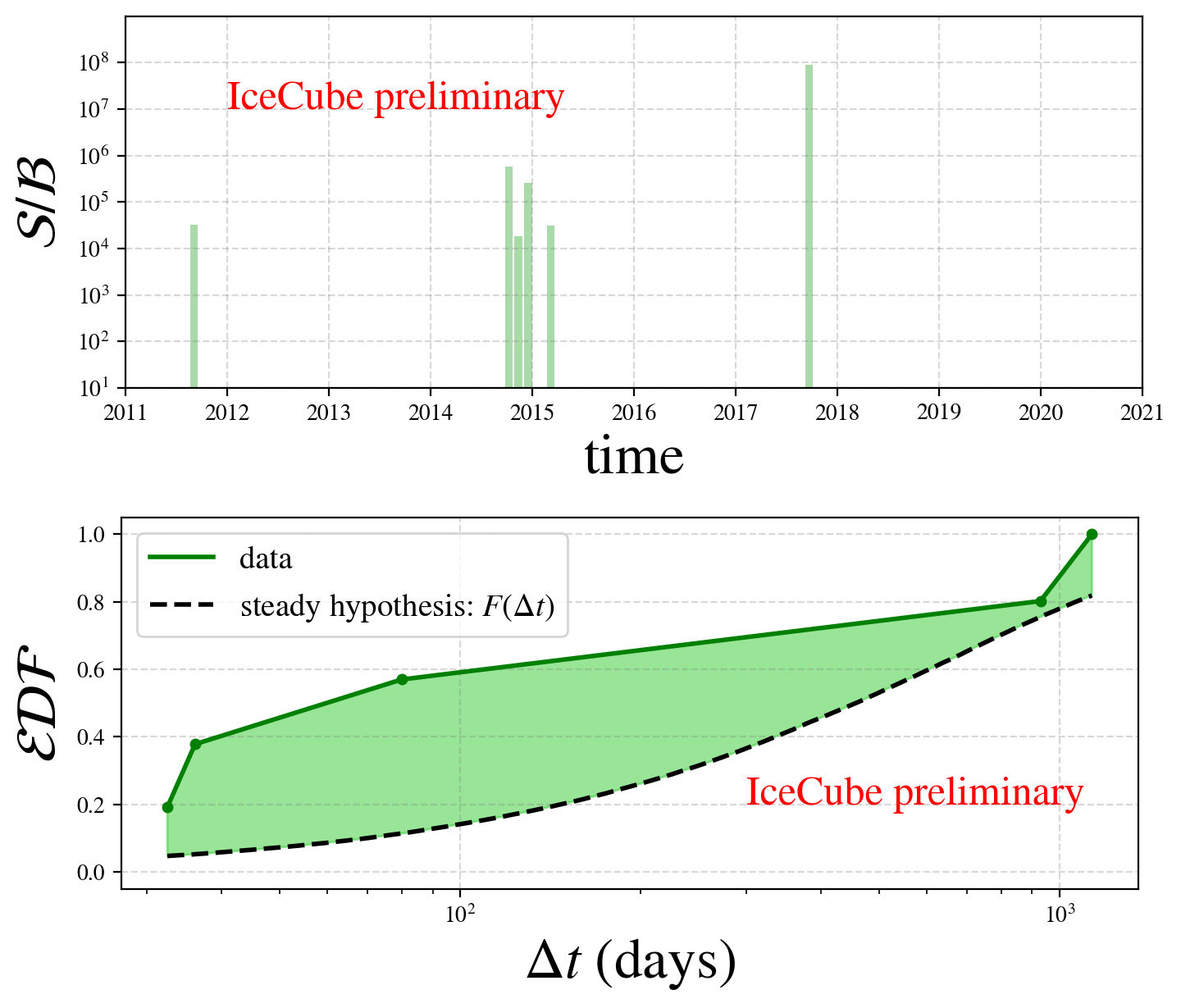}
\caption{\label{fig:txsresult} The black dashline shows the cdf for the steady hypothesis for TXS~0506+056. The green line shows the EDF.}
\end{figure}

The best fit values for signal for TXS~0506+056 are $N_{fit}=5$ and $\gamma=2.0$ and we find $N_{ev}=6$.  The top panel of figure \ref{fig:txsresult} shows the time-curve for the 6 events with the highest $\mathcal{S/B}$ for TXS~0506+056. Neutrino IceCube~170922A is clearly visible during 2017. Also visible are four events at the end of 2014 and beginning of 2015. The lower panel of figure \ref{fig:txsresult} show the CDF for the steady hypothesis, using $N_{ev}=6$, $\gamma=2.0$ and $\delta=5.7^\circ$, as we as the EDF for TXS~0506+056 data. The TAUNTON p-value for TXS~0506+056 is  0.04 ($1.7\sigma$). This is consistent with the steady hypothesis. We attribute this result to the low number of neutrino events available for TAUNTON. See that the test statistic depends on the number of signal events \ref{eq:TAUNTON}.

\section{Summary}

We presented an update to TAUNTON, a non-parametric time variability test statistic to determine compatibility of IceCube neutrino data with the steady hypothesis. We have applied TAUNTON to data used on the 4.2$\sigma$ observation of NGC~1068, notably NGC~1068 itself, as well as the time-integrated study of TXS~0506+056. We find that NGC~1068 is consistent with the steady hypothesis with a p-value of $p=0.9$. 
We find that TXS~0506+056 is also consistent with the steady hypothesis at 1.7$\sigma$. However, as discussed, this is likely the outcome of using the time-integrated analysis data for TXS~0506+056 as well as the small number of events that contribute to TXS~0506+056.

\bibliographystyle{ICRC}
\bibliography{references}

%

\clearpage

\section*{Full Author List: IceCube Collaboration}

\scriptsize
\noindent
R. Abbasi$^{17}$,
M. Ackermann$^{63}$,
J. Adams$^{18}$,
S. K. Agarwalla$^{40,\: 64}$,
J. A. Aguilar$^{12}$,
M. Ahlers$^{22}$,
J.M. Alameddine$^{23}$,
N. M. Amin$^{44}$,
K. Andeen$^{42}$,
G. Anton$^{26}$
% ...
\\
\\
% ...
$^{12}$ Universit{\'e} Libre de Bruxelles, Science Faculty CP230, B-1050 Brussels, Belgium \\
% ...
$^{17}$ Department of Physics, Loyola University Chicago, Chicago, IL 60660, USA \\
$^{18}$ Dept. of Physics and Astronomy, University of Canterbury, Private Bag 4800, Christchurch, New Zealand \\
% ...
$^{22}$ Niels Bohr Institute, University of Copenhagen, DK-2100 Copenhagen, Denmark \\
$^{23}$ Dept. of Physics, TU Dortmund University, D-44221 Dortmund, Germany \\
% ...
$^{26}$ Erlangen Centre for Astroparticle Physics, Friedrich-Alexander-Universit{\"a}t Erlangen-N{\"u}rnberg, D-91058 Erlangen, Germany \\
% ...
$^{40}$ Dept. of Physics and Wisconsin IceCube Particle Astrophysics Center, University of Wisconsin{\textendash}Madison, Madison, WI 53706, USA \\
% ...
$^{42}$ Department of Physics, Marquette University, Milwaukee, WI, 53201, USA \\
% ...
$^{44}$ Bartol Research Institute and Dept. of Physics and Astronomy, University of Delaware, Newark, DE 19716, USA \\
% ...
$^{63}$ Deutsches Elektronen-Synchrotron DESY, Platanenallee 6, 15738 Zeuthen, Germany  \\
$^{64}$ Institute of Physics, Sachivalaya Marg, Sainik School Post, Bhubaneswar 751005, India \\
% ...

\subsection*{Acknowledgements}

\noindent
The authors gratefully acknowledge the support from the following agencies and institutions:
USA {\textendash} U.S. National Science Foundation-Office of Polar Programs,
U.S. National Science Foundation-Physics Division,
U.S. National Science Foundation-EPSCoR,
Wisconsin Alumni Research Foundation,
Center for High Throughput Computing (CHTC) at the University of Wisconsin{\textendash}Madison,
Open Science Grid (OSG),
Advanced Cyberinfrastructure Coordination Ecosystem: Services {\&} Support (ACCESS),
Frontera computing project at the Texas Advanced Computing Center,
U.S. Department of Energy-National Energy Research Scientific Computing Center,
Particle astrophysics research computing center at the University of Maryland,
Institute for Cyber-Enabled Research at Michigan State University,
and Astroparticle physics computational facility at Marquette University;
Belgium {\textendash} Funds for Scientific Research (FRS-FNRS and FWO),
FWO Odysseus and Big Science programmes,
and Belgian Federal Science Policy Office (Belspo);
Germany {\textendash} Bundesministerium f{\"u}r Bildung und Forschung (BMBF),
Deutsche Forschungsgemeinschaft (DFG),
Helmholtz Alliance for Astroparticle Physics (HAP),
Initiative and Networking Fund of the Helmholtz Association,
Deutsches Elektronen Synchrotron (DESY),
and High Performance Computing cluster of the RWTH Aachen;
Sweden {\textendash} Swedish Research Council,
Swedish Polar Research Secretariat,
Swedish National Infrastructure for Computing (SNIC),
and Knut and Alice Wallenberg Foundation;
European Union {\textendash} EGI Advanced Computing for research;
Australia {\textendash} Australian Research Council;
Canada {\textendash} Natural Sciences and Engineering Research Council of Canada,
Calcul Qu{\'e}bec, Compute Ontario, Canada Foundation for Innovation, WestGrid, and Compute Canada;
Denmark {\textendash} Villum Fonden, Carlsberg Foundation, and European Commission;
New Zealand {\textendash} Marsden Fund;
Japan {\textendash} Japan Society for Promotion of Science (JSPS)
and Institute for Global Prominent Research (IGPR) of Chiba University;
Korea {\textendash} National Research Foundation of Korea (NRF);
Switzerland {\textendash} Swiss National Science Foundation (SNSF);
United Kingdom {\textendash} Department of Physics, University of Oxford.

\end{document}